\documentclass{article}
\usepackage[sc]{mathpazo} 
\usepackage[T1]{fontenc} 
\usepackage{microtype} 
\usepackage[english]{babel}
\usepackage[hmarginratio=1:1,top=25mm,left=20mm, right=30mm, columnsep=20pt]{geometry} 
\usepackage[hang, small,labelfont=bf,up,textfont=it,up]{caption} 
\usepackage{booktabs}
\usepackage{enumitem}
\setlist[itemize]{noitemsep}
\usepackage{abstract} 
\usepackage{titlesec} 
\titleformat{\section}[block]{\large\centering}{\thesection}{1em}{\MakeUppercase}{} 
\titleformat{\subsection}[block]{\large}{\thesubsection.}{1em}{} 
\usepackage{fancyhdr} 
\usepackage{titling} 
\usepackage{hyperref} 
\usepackage{array} 
\usepackage{longtable} 
\usepackage{rotating} 
\graphicspath{{./images/} }
\usepackage{subfiles} 
\usepackage[style=apa]{biblatex}
\usepackage{csquotes}
\usepackage[font=small, labelfont=normalfont, textfont=up,]{caption}

\title{DoWhy: An End-to-End Library for Causal Inference} 
\author{%
\textsc{Amit Sharma, Emre K\i c\i man}
\\[1ex] 
\textsc{Microsoft Research}
}
\date{} 
\renewcommand{\maketitlehookd}{%

\begin{abstract}
\noindent  
In addition to efficient statistical estimators of a treatment's effect, successful application of causal inference requires specifying assumptions about the  mechanisms underlying observed data and testing whether they are valid, and to what extent. However, most libraries for causal inference focus only on the task of providing powerful statistical estimators. We describe DoWhy, an open-source Python library that is built with causal assumptions as its first-class citizens, based on  the formal framework of causal graphs to specify and test causal assumptions. DoWhy presents an API for the four steps common to  any causal analysis---1) \textbf{modeling} the data using a causal graph and structural assumptions, 2) \textbf{identifying} whether the desired effect is estimable under the causal model, 3) \textbf{estimating} the effect using statistical estimators, and finally 4) \textbf{refuting} the obtained estimate through robustness checks and sensitivity analyses. In particular, DoWhy implements a number of robustness checks including placebo tests, bootstrap tests, and tests for unoberved confounding.  DoWhy is an extensible library that supports interoperability with other implementations, such as EconML and CausalML for the the estimation step. The library is available at \url{https://github.com/microsoft/dowhy}.
\end{abstract}
}

\begin{document}
\maketitle

\section{Introduction}
 Many questions in data science are fundamentally causal questions, such as the impact of a marketing campaign or a new product feature, the reasons for customer churn, which drug may work best for which patient, and so on. As the field of data science has grown, many practitioners are realizing the value of causal inference in providing insights from data. However, unlike the streamlined experience for supervised machine learning with libraries like Tensorflow~(\cite{tensorflow}) and PyTorch~(\cite{pytorch}), it is non-trivial to build a causal inference analysis. Software libraries that implement state-of-the art causal inference methods can accelerate the adoption of causal inference among data analysts in both industry and academia.

However, we find that for data scientists and machine learning engineers familiar with non-causal methods and unpracticed in the use of causal methods, one of the biggest challenges is the practice of modeling assumptions (i.e., translating domain knowledge into a causal graph) and the implications of these assumptions for causal identification and estimation.
\textit{What is the right model? }
Another challenge is in the shift in verification and testing practicalities.  Unlike supervised machine learning models that can be validated using held-out test data,  causal tasks often have no ground truth answer available.  Thus, checking core assumptions and applying sensitivity tests is critical to gaining confidence in results.
\textit{But how to check those assumptions?}

Therefore, we built DoWhy, an end-to-end library for causal analysis that builds on the latest research in modeling assumptions and robustness checks~(\cite{athey2017state,kddtutorial}), and provides an easy interface for analysts to follow the best practices of causal inference. Specifically, DoWhy's API is organized around the four key steps that are required for any causal analysis:  Model, Identify, Estimate,  and Refute. 
 \textbf{Model} encodes prior knowledge as a formal causal graph, \textbf{identify} uses graph-based methods to identify the causal effect, \textbf{estimate} uses statistical methods for estimating the identified estimand, and finally \textbf{refute} tries to refute the obtained estimate by testing robustness to initial model's assumptions.
 
The focus on all the four steps, going from data to the final causal estimate (along with a measure of its robustness) is the key differentiator for DoWhy, compared to many existing libraries for causal inference in Python and R that only focus on estimation (the third step). These libraries expect an analyst to have already figured out how to build a
reasonable causal model from data and domain knowledge, and to have identified the correct estimand. More critically, they also assume that the analyst may perform their own sensitivity and robustness checks, but provide no guidance on their own; which makes it hard to verify and build robust causal analyses.

Under the hood, DoWhy builds on two of the most powerful frameworks for causal inference: graphical models~(\cite{pearl2009causality}) and potential outcomes~(\cite{imbens2015causal}). It uses graph-based criteria and do-calculus for modeling assumptions and identifying a non-parametric causal effect. For estimation, it switches to methods based primarily on potential outcomes.  DoWhy is also built to be interoperable with other libraries that implement the estimation step. It currently supports calling EconML~(\cite{econml}) and CausalML~(\cite{causalml}) estimators. 

To summarize, DoWhy provides a unified interface for causal inference methods and automatically tests many assumptions, thus making inference accessible to non-experts. DoWhy is available open-source on Github, 
{\em \textbf{\url{https://github.com/microsoft/dowhy}}},  and has a growing community, including over 2300 stars and 31 contributors. Many people have made key contributions that are improving the usability and functionality of the library such as an integrated Pandas interface for DoWhy's four steps, and we welcome more community contributions. The library makes three key contributions: 
\begin{enumerate}
\item Provides a principled way of modeling a given problem as a causal graph so that all assumptions are explicit, and identifying a desired causal effect.
\item Provides a unified interface for many popular causal inference \textit{estimation} methods, combining the two major frameworks of graphical models and potential outcomes.
\item Automatically tests for the validity of causal assumptions if possible and assesses the robustness of the estimate to violations.
\end{enumerate}

\section{DoWhy and the four steps of causal inference}
\begin{figure}
    \centering
    \includegraphics[scale=0.5]{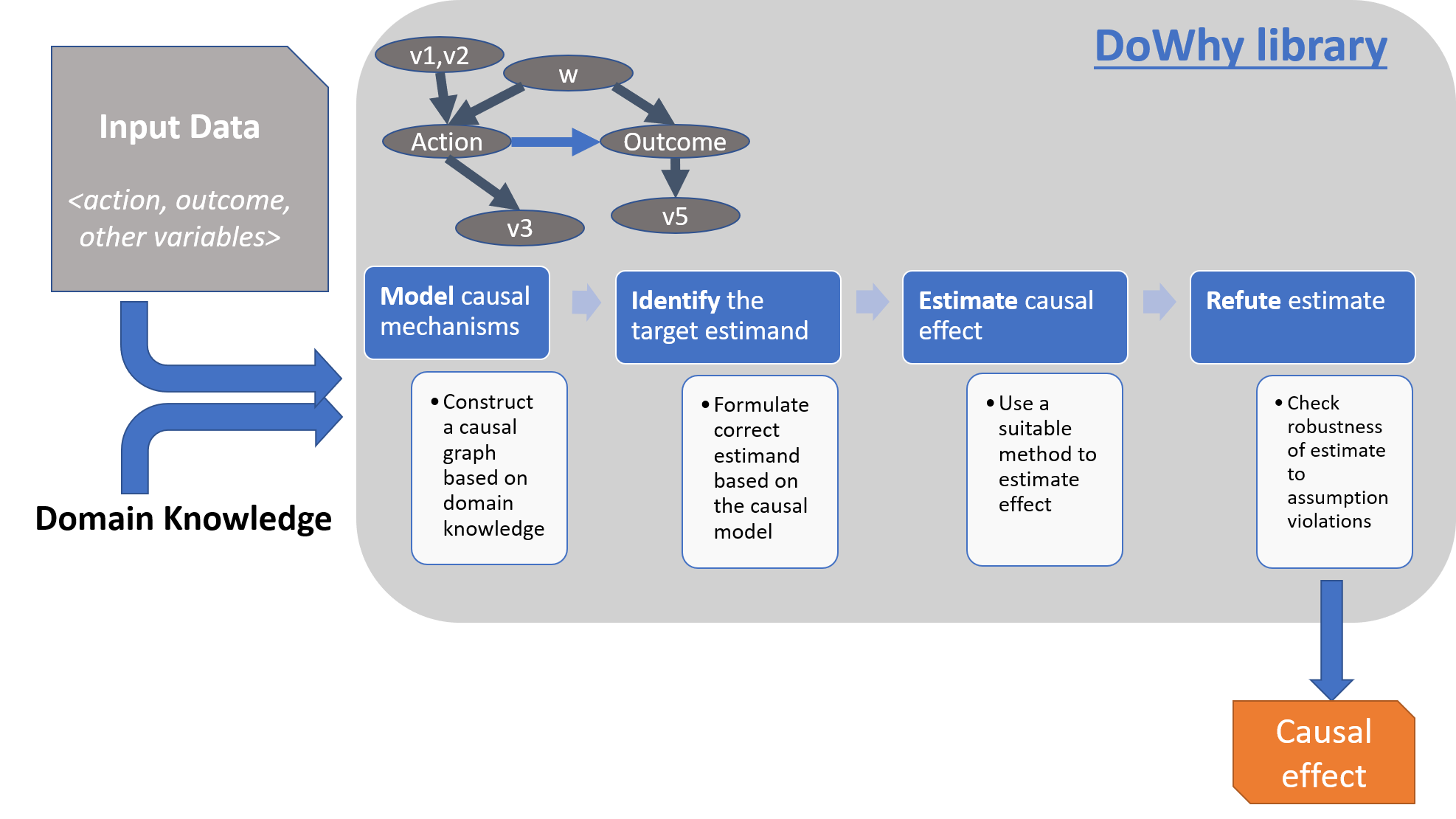}
    \caption{The four-step analysis pipeline in DoWhy. }
    \label{fig:dowhy-four-steps}
\end{figure}

DoWhy is based on a simple unifying language for causal inference. Causal inference may seem tricky, but almost all methods follow four key steps. Figure~\ref{fig:dowhy-four-steps} shows a schematic of the DoWhy analysis pipeline. 

\medskip
\noindent \textbf{I. Model the causal question.} DoWhy creates an underlying causal graphical model~(\cite{pearl2009causality}) for each problem. This serves to make each causal assumption explicit. This graph need not be complete---an analyst may  provide a partial graph, representing prior knowledge about some of the variables. DoWhy automatically considers the rest of the variables as potential confounders.

\medskip
\noindent \textbf{II. Identify the causal estimand. }Based on the causal graph, DoWhy finds all possible ways of identifying a desired causal effect based on the graphical model. It uses graph-based criteria and do-calculus to find potential ways find expressions that can identify the causal effect.

Supported identification criteria are,
\begin{itemize}
\item Back-door criterion
\item Front-door criterion
\item Instrumental Variables
\item Mediation (Direct and indirect effect identification)
\end{itemize}

\medskip
\noindent \textbf{III.  Estimate the causal effect.}
DoWhy supports methods based on both back-door criterion and instrumental variables. It also provides a non-parametric confidence intervals and a permutation test for testing the statistical significance of obtained estimate.

Supported estimation methods include, 
\begin{itemize}
\item Methods based on estimating the treatment assignment: Propensity-based Stratification, Propensity Score Matching, Inverse Propensity Weighting
\item Methods based on estimating the outcome model: Linear Regression, Generalized Linear Models
\item Methods based on the instrumental variables identification:
Binary Instrument/Wald Estimator, Two-stage least squares, Regression discontinuity
\item Methods for front-door criterion and general mediation: 
Two-stage linear regression
\end{itemize}

In addition, DoWhy support integrations with the EconML and CausalML packages for estimating the conditional average treatment effect (CATE). All estimators from these libraries can be  directly called from DoWhy. 

\medskip
\noindent \textbf{IV. Refute the obtained estimate.}
Having access to multiple refutation methods to validate an effect estimate from a causal estimator is a key benefit of using DoWhy.

Supported refutation methods include:
\begin{itemize}
\item \textbf{Add Random Common Cause:} Does the estimation method change its estimate after we add an independent random variable as a common cause to the dataset? (Hint: It should not)
\item \textbf{Placebo Treatment}: What happens to the estimated causal effect when we replace the true treatment variable with an independent random variable? (Hint: the effect should go to zero)
\item \textbf{Dummy Outcome:} What happens to the estimated causal effect when we replace the true outcome variable with an independent random variable? (Hint: The effect should go to zero)
\item \textbf{Simulated Outcome:} What happens to the estimated causal effect when we replace the outcome with a simulated outcome based on a known data-generating process closest to the given dataset? (Hint: It should match the effect parameter from the data-generating process)
\item \textbf{Add Unobserved Common Causes: }How sensitive is the effect estimate when we add an additional common cause (confounder) to the dataset that is correlated with the treatment and the outcome? (Hint: It should not be too sensitive)
\item \textbf{Data Subsets Validation:} Does the estimated effect change significantly when we replace the given dataset with a randomly selected subset? (Hint: It should not)
\item \textbf{Bootstrap Validation}: Does the estimated effect change significantly when we replace the given dataset with bootstrapped samples from the same dataset? (Hint: It should not)
\end{itemize}

Many of the above methods aim to refute the full causal analysis, including modeling, identification and estimation (as in Placebo Treatment or Dummy Outcome) whereas others refute a specific step (e.g., Data Subsets and Bootstrap Validation that test only the estimation step). 

\section{An example causal analysis}
In this section, we show how causal inference using DoWhy simplifies to four lines of code, each corresponding to one of the four steps. Each analysis starts with a building a causal model. The assumptions can be viewed graphically or in terms of conditional independence statements. Wherever possible, DoWhy can also automatically test for stated assumptions using observed data.

\begin{verbatim}
# I. Create a causal model from the data and given graph.
model = CausalModel(
    data=data["df"],
    treatment=data["treatment_name"],
    outcome=data["outcome_name"],
    graph=data["gml_graph"])
\end{verbatim}
Given the model, identification is a causal problem. Estimation is simply a statistical problem. 
DoWhy respects this boundary and treats them separately. This focuses the causal inference effort on identification, and frees up estimation using any available statistical estimator for a target estimand. In addition, multiple estimation methods can be used for a single identified estimand and vice-versa.

\begin{verbatim}
# II. Identify causal effect and return target estimands
identified_estimand = model.identify_effect()

# III. Estimate the target estimand using a statistical method.
estimate = model.estimate_effect(identified_estimand,
                                 method_name="backdoor.propensity_score_stratification")
\end{verbatim}
For data with high-dimensional confounders, machine learning-based estimators may be more effective. Therefore, DoWhy supports calling estimators from other libraries like EconML. Here is an example of using the double machine learning estimator~(\cite{chernozhukov2017double}). 

\begin{verbatim}
import econml
dml_estimate = model.estimate_effect(
                identified_estimand, method_name="backdoor.econml.dml.DMLCateEstimator",
                confidence_intervals=False,
                method_params={
                    "init_params":{
                        `model_y':GradientBoostingRegressor(),
                        `model_t': GradientBoostingRegressor(),
                        `model_final':LassoCV(), 
                        `featurizer':PolynomialFeatures(degree=1, include_bias=True)},
                    "fit_params":{}})
print(dml_estimate)
\end{verbatim}

The most critical, and often skipped, part of causal analysis is checking the robustness of an estimate to unverified assumptions. DoWhy makes it easy to automatically run sensitivity and robustness checks on the obtained estimate.

\begin{verbatim}
# IV. Refute the obtained estimate using multiple robustness checks.
refute_results = model.refute_estimate(identified_estimand, estimate,
                                       method_name="random_common_cause")
\end{verbatim}
Finally, DoWhy is easily extensible, allowing other implementations of the four verbs to co-exist. The four verbs are mutually independent, so their implementations can be combined in any way. Example notebooks of using DoWhy for different causal problems are available at \url{https://github.com/microsoft/dowhy/tree/master/docs/source/example_notebooks}.

\section{Conclusion}
We presented DoWhy, an extensible and end-to-end library for causal inference. Unlike most other libraries, DoWhy focuses on helping an analyst devise the correct causal model and test its assumptions, in addition to estimating the causal effect. We look forward to extending DoWhy with more refutation and robustness analyses, and supporting more estimation methods with its 4-step API. 

\section*{Acknowledgements}
A big thanks to all the open-source contributors to DoWhy that continue to make important additions to the library's functionality and usability. The list of contributors is updated at \url{https://github.com/microsoft/dowhy/blob/master/CONTRIBUTING.md}. 

\printbibliography

\end{document}